**Стрюк Андрій Миколайович**
кандидат педагогічних наук, докторант
Інститут інформаційних технологій і засобів навчання НАПН України, м. Київ, Україна
*andrey.n.stryuk@gmail.com*

**Семеріков Сергій Олексійович**
професор, доктор педагогічних наук, завідувач кафедри фундаментальних і
соціально-гуманітарних дисциплін
ДВНЗ «Криворізький національний університет», м. Кривий Ріг, Україна
*semerikov@gmail.com*

**Тарасов Ігор Володимирович**
асистент кафедри інформатики та прикладної математики
ДВНЗ «Криворізький національний університет», м. Кривий Ріг, Україна
*taras_2001@rambler.ru*

# КОМПЕТЕНТНІСТЬ БАКАЛАВРА ІНФОРМАТИКИ З ПРОГРАМУВАННЯ

**Анотація.** На основі аналізу підходів до визначення професійних компетентностей фахівців з інформаційних технологій виокремлено компетентність бакалавра інформатики з програмування. З урахуванням галузевого стандарту вищої освіти з напряму підготовки 040302 «Інформатика» та Computing Curricula 2001 визначено зміст і структуру компетентності бакалавра інформатики з програмування. Спроектовано систему змістових модулів, що забезпечують її формування. Визначено внесок нормативних компетенцій бакалавра інформатики у формування компетентності з програмування. Запропоновано напрями формування компетентності з програмування, зокрема, у хмаро орієнтованому середовищі навчання.

**Ключові слова:** компетентність з програмування; підготовка бакалаврів інформатики; професійні компетенції бакалавра інформатики.

## 1. ВСТУП

**Постановка проблеми.** Однією з проблем організації навчання програмування є вибір його апаратно-програмного забезпечення, що, з одного боку, має відповідати вимогам збільшення продуктивності й надійності за постійного збільшення обсягів даних для опрацювання, а з іншого, — скорочення витрат на підтримку і розвиток ІКТ-інфраструктури і підвищення її адаптивності до потреб навчально-виховного процесу.

Одним із засобів задоволення вказаних вимог є перебудова процесу навчання на основі застосування хмарних технологій, перспективи використання яких в освітніх цілях окреслюють В. Ю. Биков [22], І. С. Войтович [6], О. С. Воронкін [7], З. М. Гадецька [8], М. І. Жалдак [11], О. О. Жугастров [12], В. П. Сергієнко [19], А. В. Колесников [13], О. Г. Кузьминська [14], Н. В. Морзе [16], З. С. Сейдаметова [17], С. Н. Сейтвелієва [18], Л. Е. Соколова [20], О. М. Спірін [21], Ю. В. Триус [23], М. А. Шиненко [24], М. П. Шишкіна [25] та інші науковці. Ці технології надають можливість використовувати зовнішні, розташовані за межами персональних комп'ютерів, високопродуктивні обчислювальні ресурси, щоб виконувати різні внутрішні завдання.

Використання хмарних технологій дозволяє навчальним закладам отримати низку переваг; зокрема, еластичність, доступність та мобільність навчання створюють умови для організації мобільного і комбінованого навчання і є технологічною основою фундаменталізації навчання програмування. Як зазначає І. С. Мінтій, навчання програмування надає можливість природно відобразити зв'язок двох головних складових інформатики: математичної інформатики та інформаційних технологій [15, с. 6], тому й компетентності з програмування посідають важливе місце в системі інформатичних компетентностей бакалавра інформатики.

**Аналіз останніх досліджень і публікацій.** Згідно документа [4] інформатичні компетентності діляться на чотири групи: інформаційні служби, мережні системи, програмування та розробка програмного забезпечення, засоби взаємодії. Основну частину документа [4] складають 49 модулів, які, зокрема, містять компетентності та їх складові з програмування: теорія програмування, мови прикладного програмування, розробка програмного забезпечення.

Департаментом освіти США визначено рекомендовану структуру компетентності з програмування:

– уміння пояснити призначення і функції комп'ютерних програм;
– уміння пояснити термін «мови програмування» і навести приклади для кожної з різних парадигм програмування;
– уміння пояснити фактори, які необхідно розглянути під час вибору мови програмування для розв'язування задачі;
– уміння описати етапи розробки програм;
– уміння пояснити і застосувати концепції програмування й інструментальні засоби, що використовуються в структурному програмуванні;
– уміння пояснити і проілюструвати лінійні, з розгалуженням та циклічні конструкції, що використовуються в структурному програмуванні;
– уміння пояснити відмінності використання різних парадигм програмування: подіє-орієнтованого програмування, об'єктно-орієнтованого програмування та імперативного програмування;
– уміння створити внутрішню і зовнішню документацію до програми;
– уміння спроектувати, написати, перевірити та дослідити результати виконання програм [1, с. 47].

Перелічені елементи зручно використати для діагностування сформованості праксеологічної складової компетентності, проте засоби діагностики інших її складових у [1] не наведено.

І. С. Мінтій у структурі спеціальних професійних компетентностей учителя інформатики виокремлює компетентності: з теоретичної (математичної) інформатики; з програмування; з інформаційних технологій; з фундаментальних природничо-математичних дисциплін [15, с. 50]. Дослідник наголошує, що набуття компетентностей з математичної інформатики приводить до якісних змін у рівні компетентностей з програмування, що, у свою чергу, приводить до змін у компетентностях з інформаційних технологій, і навпаки. Отже, компетентності з програмування відображають зв'язок між компетентностями з математичної інформатики й інформаційних технологій. Компетентності з фундаментальних природничо-математичних дисциплін мають взаємозв'язок з усіма іншими інформатичними компетентностями.

Для майбутніх учителів інформатики І. С. Мінтій запропоновано таку структуру компетентності з програмування [15, с. 52–53]:

– *когнітивно-змістову* (гносеологічну) — знання основних форм для керування виконанням програми; знання простих типів даних і функцій для роботи з ними; знання похідних типів даних, способів їх утворення з простих типів даних, функцій для роботи з ними й пріоритетних напрямів їх використання; знання основних етапів розв'язування прикладних задач; знання основних етапів проектування програм; знання складових мови програмування;

– *операційно-технологічну* (праксеологічну) — уміння пояснити призначення і

функції існуючої програми, знайти помилки в логіці розв'язання задачі, описати етапи розробки програм, розробити функції й обґрунтувати пріоритетність використання того чи іншого виразу для їх створення, створити документацію до програми, пояснити і продемонструвати процес створення похідних типів даних, спроектувати, описати, перевірити та проаналізувати результати виконання програми; оцінити переваги різних способів розв'язування однієї задачі; уміння обирати засоби для розв'язання задачі й обґрунтовувати свій вибір; уміння використовувати можливості обраних засобів (довідка, налагодження програми, налаштування необхідних параметрів та ін.);

– *ціннісно-мотиваційну* (*аксіологічну*) — емоційно-ціннісне ставлення до процесу розробки, опису, налагодження, тестування програм та аналізу результатів їх роботи; уміння знаходити нові, нестандартні рішення задачі; внутрішня мотивація до опанування програмування; готовність до активного застосування гносеологічних і праксеологічних складових у практичній діяльності; прагнення до самовдосконалення, потреба у саморозвитку гносеологічних і праксеологічних складових; уміння самостійно приймати рішення, критично ставитись до чужих впливів, здатності за власним почином організовувати діяльність, ставити мету, у разі необхідності вносити в поведінку зміни; уміння постійно і тривало домагатися мети; наполегливість у досягненні мети, прагнення до поліпшення отриманих результатів, незадоволеність досягнутим, намагання домогтися успіху; внутрішня потреба у створенні програмних продуктів;

– *соціально-поведінкову* – здатність до співпраці у процесі розробки, опису, налагодження, тестування програм та аналізу результатів їх роботи, використання засобів для організації спільної роботи над проектом; відповідальність за власну поведінку, за виконання завдань; комунікабельність; здатність до адаптації; схильність до дискусії.

Проте відкритим залишається питання структури компетентності з програмування бакалавра інформатики — майбутнього фахівця з інформаційних технологій, що зумовило визначення **мети статті**: теоретичне обґрунтування структури і змісту компетентності бакалавра інформатики з програмування.

## 2. РЕЗУЛЬТАТИ ДОСЛІДЖЕННЯ

Підготовка бакалаврів інформатики у ВНЗ України виконується у межах галузі знань «Системні науки та кібернетика». Відповідні складові галузевого стандарту вищої освіти України (ГСВО) [9; 10] затверджено Наказом МОН України №880 від 16 вересня 2010 року. Відповідно до освітньо-кваліфікаційної характеристики бакалавра інформатики, випускники бакалаврату мають подвійну кваліфікацію — «фахівець з інформаційних технологій» і «викладач-стажист» з узагальненим об'єктом діяльності — «процеси обробки інформації алгоритмічними методами з використанням комп'ютерної техніки, навчання інформатики в навчальних закладах I–II рівня акредитації» [9, с. 7].

Випускник-бакалавр з напряму підготовки «Інформатика» повинен володіти:
– знаннями, уміннями і навичками, які необхідні для розробки, упровадження і використання систем обробки інформації алгоритмічними методами з використанням комп'ютерної техніки, математичних методів і алгоритмів у різних галузях науки і народного господарства;
– основними положеннями філософії, історії України та історії культури України, володіти діловою українською й іноземною мовами, основами педагогіки;
– основними поняттями, концепціями і фактами інформатики й математики [9, с. 16].

Підготовка бакалавра з інформатики передбачає його готовність працювати й набувати навички знань з інформаційних технологій, математичного і комп'ютерного моделювання процесів і систем різної природи, задач прогнозування, оптимізації, системного аналізу та прийняття рішень тощо [9, с. 17].

Розробники ГСВО зазначають, що «... в Україні чітко розмежовано підготовку спеціалістів з інформатики у галузях, які мають фундаментальне та практичне спрямування. Ця обставина певним чином полегшує задачу розробки стандарту для галузі знань 0403, бо дозволяє надати пріоритет фундаментальній складовій освіти. Тим самим, буде продовжена традиція радянської освіти, яка орієнтувалась на високий рівень підготовки випускників з інформатики та прикладної математики» [5, с. 9]. Отже, підготовка бакалаврів інформатики має насамперед фундаментальну спрямованість, у той час як підготовка бакалаврів програмної інженерії — насамперед прикладну.

Автори галузевого стандарту вищої освіти з напряму підготовки 040302 «Інформатика» у списку рекомендованих джерел наводять посилання на Computing Curricula 2001 (CC2001) [3], оновлена версія якого [2] містить ядро знань: DS (дискретні структури), PF (основи програмування), AL (теорія алгоритмів), AR (архітектура комп'ютерних систем), OS (операційні системи), NC (розподілені обчислення), PL (мови програмування), HC (людино-машинний інтерфейс), GV (графіка і візуалізація), IS (інтелектуальні системи), IM (інформаційний менеджмент), SP (соціальні і професійні питання), SE (програмна інженерія), CN (обчислювальна математика і чисельні методи). На рис. 1 показано внесок блоку програмування (модулі PF та PL) у загальний навчальний план підготовки згідно CC2001.

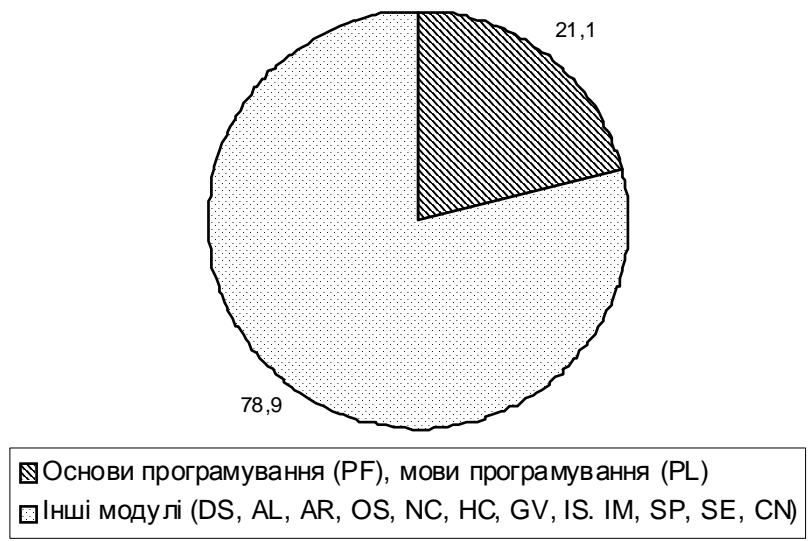

*Рис. 1. Місце програмування в інформатичній підготовці за CC2001*

До основних компетенцій, що визначаються освітньо-кваліфікаційною характеристикою бакалавра інформатики, належать такі: соціально-особистісні (КСО.01–08), загальнонаукові (КЗН.01–05), інструментальні (КІ.01–06), загально-професійні (КЗП.01–07) та спеціалізовано-професійні (КСП.01–18). Серед загально-професійних компетенцій до програмування відноситься насамперед КЗП-5 (знання та розуміння основ програмування, мов різних рівнів та їхніх переваг для розв'язання конкретних задач, методів розроблення програмного забезпечення комп'ютеризованих систем з використанням сучасних технологій).

Виробничі функції, якими повинні володіти бакалаври інформатики: *дослідницька*

(спрямована на збір, обробку, аналіз і систематизацію науково-технічної інформації з напрямку роботи); *контрольна* (спрямована на здійснення контролю в межах своєї професійної діяльності в обсязі посадових обов'язків), *проектувальна* (*проектувально-конструкторська*) — функція спрямована на здійснення цілеспрямованої послідовності дій щодо синтезу систем або окремих їх складових, розробку документації, яка необхідна для втілення і використання об'єктів і процесів, *прогностична* (функція, яка дозволяє на основі аналізу здійснювати прогнозування в професійній діяльності), *організаційна* (спрямована на упорядкування структури й взаємодії складових елементів системи з метою зниження невизначеності, а також підвищення ефективності використання ресурсів і часу), *управлінська* (спрямована на досягнення поставленої мети, забезпечення сталого функціонування і розвитку систем завдяки інформаційному обміну), *технологічна* (спрямована на втілення поставленої мети за відомими алгоритмами), *технічна* (спрямована на виконання технічних робіт у професійній діяльності) [9, с. 12].

У табл. 1 наведено частину переліку нормативних навчальних дисциплін і практик бакалавра інформатики, що відповідають стандарту CC2001 [10]. Слід відзначити, що навчальна дисципліна «Теорія програмування» відповідає тим змістовим модулям модуля PL (мови програмування), що за стандартом CC2001 не є обов'язковим для вивчення, у той час як за [10] ця навчальна дисципліна відноситься до нормативних.

*Таблиця 1*

**Навчальні дисципліни і змістові модулі підготовки бакалаврів інформатики з програмування**

| Назва навчальної дисципліни | Назва блоку змістових модулів, що входить до навчальної дисципліни | Назва змістового модуля |
|---|---|---|
| Алгоритми і структури даних | Теорія складності алгоритмів | Побудова та аналіз алгоритмів. Класифікація алгоритмів. Абстрактні типи даних |
| | | Оцінки складності алгоритмів, класифікація алгоритмів за складністю. NP-повнота алгоритмів |
| | Основні структури даних та алгоритми їх обробки | Вступ в структури даних. Класифікація структур даних (масиви, записи, черги, стеки, лінійні списки, текстові файли тощо). Динамічний розподіл пам'яті |
| | | Алгоритми сортування та пошуку |
| | Табличні структури даних, графи, дерева та алгоритми їх обробки | Дерева, як структури даних та алгоритми їх обробки |
| | | Табличні структури даних та алгоритми їх обробки |
| | | Графи, як структури даних та алгоритми їх обробки |
| Програмування | Основи програмування алгоритмічною мовою | Інструменти і базові засоби програмування |
| | | Команди та дані. Структури керування |
| | Основні концепції | Абстракція даних. Складені структури |

| Назва навчальної дисципліни | Назва блоку змістових модулів, що входить до навчальної дисципліни | Назва змістового модуля |
|---|---|---|
| | алгоритмічних мов | даних |
| | | Алгоритмічна декомпозиція |
| | Процедурне програмування | Базові засоби процедурного програмування C/C++ |
| | | Особливості процедурного програмування на базі C/C++ |
| | Об'єкто-зорієнтоване програмування | Об'єктне програмування |
| | | Ієрархічне програмування |
| Теорія програмування | Основні поняття теорії програмування | Основні аспекти програм. Розвиток та уточнення основних понять програмування |
| | | Природні та формальні мови. Підходи до формалізації мов специфікацій та програмування |
| | Синтактика: формальні мови та граматики | Методи подання синтаксису мов програмування. Формальні мови та граматики |
| | | Автоматні формалізми сприйняття мов. Розв'язні та нерозв'язні проблеми теорії формальних мов |
| | Семантика програм | Методи подання семантики програм |
| | | Рекурсія в мовах програмування. Теорія найменшої нерухомої точки та її застосування |
| | | Методи аналізу, верифікації та формальної розробки програм |

Згідно освітньо-професійної програми підготовки бакалавра інформатики, державна атестація має три основні нормативні форми: курсові роботи, дипломна робота, державний іспит.

Тематика курсових робіт встановлюється відповідної до освітньо-кваліфікаційних характеристик бакалаврів за напрямом підготовки 040302 «Інформатика» і системи з 11 змістових модулів, 3 з яких («Теорія складності алгоритмів», «Основні структури даних та алгоритми їх обробки», «Табличні структури даних, графи, дерева та алгоритми їх обробки») відносяться до навчальної дисципліни «Алгоритми та структури даних». Ці ж модулі забезпечують і виконання випускної дипломної роботи. Єдиний додатковий змістовий модуль, що виноситься на державний іспит — «Об'єкто-зорієнтоване програмування» — відповідає об'єктно-орієнтованій технології програмування. У державній атестації бакалавра інформатики в Україні змістові модулі з програмування займають 23,5% (проти 21,1% у міжнародному стандарті CC2001).

Наведені у табл. 1 змістові модулі з програмування забезпечують формування виробничих функцій і вмінь бакалавра інформатики (табл. 2).

*Таблиця 2*

**Виробничі функції, типові задачі діяльності та уміння бакалавра інформатики, що формуються у процесі навчання програмування**

| Типова задачі діяльності | Зміст уміння, що забезпечується |
|---|---|
| *Дослідницька виробнича функція* | |
| Розробка математично обґрунтованих алгоритмів функціонування комп'ютеризованих систем | Вміти використовувати, розробляти та досліджувати математичні методи та алгоритми обробки даних (статистичні, алгебраїчні, комбінаторні, теоретико-інформаційні та інші). *Змістовий модуль:* Побудова та аналіз алгоритмів. Класифікація алгоритмів. Абстрактні типи даних |
| | Вміти використовувати, розробляти та досліджувати алгоритми розв'язування задач моделювання об'єктів і процесів інформатизації, задач оптимізації, прогнозування, оптимального керування та прийняття рішень, тощо. *Змістові модулі:* 1. Побудова та аналіз алгоритмів. Класифікація алгоритмів. Абстрактні типи даних. 2. Оцінки складності алгоритмів, класифікація алгоритмів за складністю. NP-повнота алгоритмів. 3. Вступ до структур даних. Класифікація структур даних (масиви, записи, черги, стеки, лінійні списки, текстові файли тощо). Динамічний розподіл пам'яті. 4. Алгоритми сортування та пошуку. 5. Дерева, як структури даних та алгоритми їх обробки. 6. Табличні структури даних та алгоритми їх обробки. 7. Графи, як структури даних та алгоритми їх обробки |
| | Вміти використовувати, розробляти та досліджувати алгоритми функціонування комп'ютеризованих систем методами неперервної, дискретної математики, математичної логіки тощо. *Змістовий модуль:* Оцінки складності алгоритмів, класифікація алгоритмів за складністю. NP-повнота алгоритмів |
| | Вміти оцінювати складові ефективності алгоритмів функціонування комп'ютеризованих систем. *Змістові модулі:* 1. Побудова та аналіз алгоритмів. Класифікація алгоритмів. Абстрактні типи даних. 2. Оцінки складності алгоритмів, класифікація алгоритмів за складністю. NP-повнота алгоритмів |
| *Проектувальна виробнича функція* | |
| Проектування програмного забезпечення комп'ютеризованих систем | Вміти використовувати основні парадигми проектування програмного забезпечення: структурну, об'єкто-зорієнтовану, компонентну, аспектно-орієнтовану, сервіс-орієнтовану, мультиагентну, розподілену тощо для розробки проекту комп'ютеризованої системи. *Змістові модулі:* 1. Базові засоби процедурного програмування C/C++. 2. Об'єктне програмування. 3. Ієрархічне програмування |
| | Володіти методами опису основних понять |

| Типова задачі діяльності | Зміст уміння, що забезпечується |
|---|---|
| | програмування, вміти задавати семантику та синтаксис конструкцій мов програмування.<br>*Змістові модулі:*<br>1. Основні аспекти програм. Розвиток та уточнення основних понять програмування.<br>2. Природні та формальні мови. Підходи до формалізації мов специфікацій та програмування.<br>3. Методи подання синтаксису мов програмування. Формальні мови та граматики.<br>4. Автоматні формалізми сприйняття мов. Розв'язні та нерозв'язні проблеми теорії формальних мов.<br>5. Методи подання семантики програм.<br>6. Рекурсія в мовах програмування. Теорія найменшої нерухомої точки та її застосування.<br>7. Методи аналізу, верифікації та формальної розробки програм |
| *Технологічна виробнича функція* | |
| Використання програмного забезпечення комп'ютеризованих систем | Володіти основами програмування та мовами різних рівнів (машинними, асемблерними, високого рівня, проблемно та предметно орієнтованими).<br>*Змістові модулі:*<br>1. Інструменти і базові засоби програмування.<br>2. Команди та дані. Структури керування.<br>3. Абстракція даних. Складені структури даних.<br>4. Алгоритмічна декомпозиція.<br>5. Базові засоби процедурного програмування С/С++.<br>6. Особливості процедурного програмування на базі С/С++.<br>7. Об'єктне програмування.<br>8. Ієрархічне програмування |
| Технології розроблення програмного забезпечення комп'ютеризованих систем | Вміти розробляти програмне забезпечення комп'ютеризованої системи з використанням технологій програмування, заснованими на структурній, об'єкто-зорієнтованій, компонентній, аспектно-орієнтованій, сервіс-орієнтованій, мультиагентній, розподіленій, логічній та інших парадигмах.<br>*Змістові модулі:*<br>1. Основні аспекти програм. Розвиток та уточнення основних понять програмування.<br>2. Природні та формальні мови. Підходи до формалізації мов специфікацій та програмування.<br>3. Методи подання синтаксису мов програмування. Формальні мови та граматики.<br>4. Автоматні формалізми сприйняття мов. Розв'язні та нерозв'язні проблеми теорії формальних мов.<br>5. Методи подання семантики програм.<br>6. Рекурсія в мовах програмування. Теорія найменшої нерухомої точки та її застосування.<br>7. Методи аналізу, верифікації та формальної розробки |

| Типова задачі діяльності | Зміст уміння, що забезпечується |
|---|---|
| | програм.<br>8. Інструменти і базові засоби програмування.<br>9. Команди та дані. Структури керування.<br>10. Абстракція даних. Складені структури даних.<br>11. Алгоритмічна декомпозиція.<br>12. Базові засоби процедурного програмування C/C++.<br>13. Особливості процедурного програмування на базі C/C++.<br>14. Об'єктне програмування.<br>15. Ієрархічне програмування |
| *Контрольна виробнича функція* | |
| Контроль за виконанням робіт з розроблення комп'ютеризованих систем | Вміти контролювати правильність роботи програмного забезпечення розробленої комп'ютеризованої системи за допомогою тестування на різних рівнях (модульному, інтеграційному, системному тощо).<br>*Змістовий модуль:* Методи аналізу, верифікації та формальної розробки програм |

Компетентність — інтегрована характеристика якостей особистості, результат підготовки випускника ВНЗ для виконання діяльності в певних професійних і соціально-особистістних предметних галузях (компетенціях), який визначається необхідним обсягом і рівнем знань і досвіду в певному виді діяльності [9, с. 13]. Враховуючи [9, с. 15], що кожній типовій задачі діяльності бакалавра інформатики відповідає компетенція, яка формується системою умінь щодо розв'язання цієї задачі діяльності, покажемо, які компетенції забезпечують діяльність бакалавра інформатики з програмування.

Першу типову задачу діяльності бакалавра інформатики з програмування *«Розробка математично обґрунтованих алгоритмів функціонування комп'ютеризованих систем»* забезпечують компетенції:

КІ-1. Знання та розуміння правил письмової й усної рідної мови;

КІ-2. Знання та розуміння правил письмової й усної іноземної мови (мов);

КІ-6. Знання законів, методів та методик проведення наукових та прикладних досліджень;

КСП-1. Знання та розуміння методів системного аналізу та теоретичної кібернетики щодо побудови інформаційних моделей об'єктів та процесів різної природи;

КЗН-4. Базові знання із системних та кібернетичних наук, необхідних для засвоєння загально-професійних дисциплін з інформатики;

КЗП-1. Знання методології системних досліджень, методів дослідження та аналізу складних природних, техногенних, економічних та соціальних об'єктів та процесів, розуміння складності об'єктів та процесів різної природи, їх різноманіття, багатофункціональність, взаємодію та умови існування для розв'язання прикладних і наукових завдань в галузі системних наук та кібернетики.

Другу типову задачу діяльності бакалавра інформатики з програмування *«Проектування програмного забезпечення комп'ютеризованих систем»* забезпечують компетенції:

КІ-5. Знання методів та правил роботи з комп'ютером та роботи в Інтернеті;

КІ-7. Володіння навичками користувача офісних технологій у контексті опрацювання економічної інформації;

КСП-7. Знання основних парадигм проектування та мов моделювання програмного забезпечення комп'ютеризованих систем, методів планування життєвого циклу програмного забезпечення та розроблення моделі керування ресурсами;

КЗН-4. Базові знання із системних та кібернетичних наук, необхідних для засвоєння загально-професійних дисциплін з інформатики;

КЗН-5. Базові знання в галузі інформатики й сучасних інформаційних технологій;

КЗП-4. Знання вимог чинних державних та міжнародних стандартів, методів і засобів проектування комп'ютеризованих систем, життєвого циклу їх програмного забезпечення.

Третю типову задачу діяльності бакалавра інформатики з програмування *«Використання програмного забезпечення комп'ютеризованих систем»* забезпечують компетенції:

КСП-15. Знання операційних систем (Windows, Unix тощо), системного програмного забезпечення, найбільш розповсюджених пакетів прикладних програм, інформаційних порталів Інтернет, програмних методів захисту інформації в комп'ютеризованих системах та мережах;

КЗН-5. Базові знання в галузі інформатики й сучасних інформаційних технологій;

КІ-5. Знання методів та правил роботи з комп'ютером та роботи в Інтернеті;

КІ-6. Знання законів, методів та методик проведення наукових та прикладних досліджень;

КЗП-5. Знання та розуміння основ програмування, мов різних рівнів та їхніх переваг для розв'язання конкретних задач, методів розроблення програмного забезпечення комп'ютеризованих систем з використанням сучасних технологій;

КСП-17. Знання методів, методик контролю та тестування правильності роботи програмного забезпечення комп'ютеризованих систем.

Четверту типову задачу діяльності бакалавра інформатики з програмування *«Технології розроблення програмного забезпечення комп'ютеризованих систем»* забезпечують компетенції:

КСП-16. Знання базових та спеціалізованих технологій розроблення програмного забезпечення комп'ютеризованих систем;

КЗН-5. Базові знання в галузі інформатики й сучасних інформаційних технологій;

КІ-5. Знання методів та правил роботи з комп'ютером та роботи в Інтернеті;

КІ-6. Знання законів, методів та методик проведення наукових та прикладних досліджень;

КЗП-5. Знання та розуміння основ програмування, мов різних рівнів та їхніх переваг для розв'язання конкретних задач, методів розроблення програмного забезпечення комп'ютеризованих систем з використанням сучасних технологій.

П'яту типову задачу діяльності бакалавра інформатики з програмування *«Контроль за виконанням робіт з розроблення комп'ютеризованих систем»* забезпечують компетенції:

КСО-7. Знання та розуміння законів та методів міжособистісних комунікацій, норм толерантності, ділових комунікацій у професійній сфері, ефективної праці в колективі, адаптивності;

КЗП-4. Знання вимог чинних державних та міжнародних стандартів, методів і засобів проектування комп'ютеризованих систем, життєвого циклу їх програмного забезпечення;

КЗП-7. Знання основних методів та підходів щодо організації, планування, керування та контролю роботами з проектування, розроблення, післяпроектного супроводу та експлуатації програмного забезпечення комп'ютеризованих систем;

КСП-13. Знання методів, нормативів, державних стандартів та чинного законодавства стосовно організації, планування, контролю та управління роботами з проектування та розроблення комп'ютеризованих систем колективом розробників;

КСП-17. Знання методів, методик контролю та тестування правильності роботи програмного забезпечення комп'ютеризованих систем.

Перелічені компетенції щодо розв'язання типових задач діяльності бакалавра інформатики з програмування є нормативною основою *компетентності бакалавра інформатики з програмування*, структуру якої подано на рис. 2. Шифри компетенцій виділено напівжирним шрифтом. Шифри, унікальні для певної типової задачі діяльності бакалавра інформатики з програмування, подано у прямокутниках, а шифри компетенцій, спільні для кількох задач діяльності, подано в овалах.

Із рис. 2 видно, що компетенції КЗН-5 (базові знання в галузі інформатики й сучасних інформаційних технологій), КІ-5 (знання методів та правил роботи з комп'ютером та роботи в Інтернеті) та КІ-6 (знання законів, методів та методик проведення наукових та прикладних досліджень) є спільними для трьох видів діяльності.

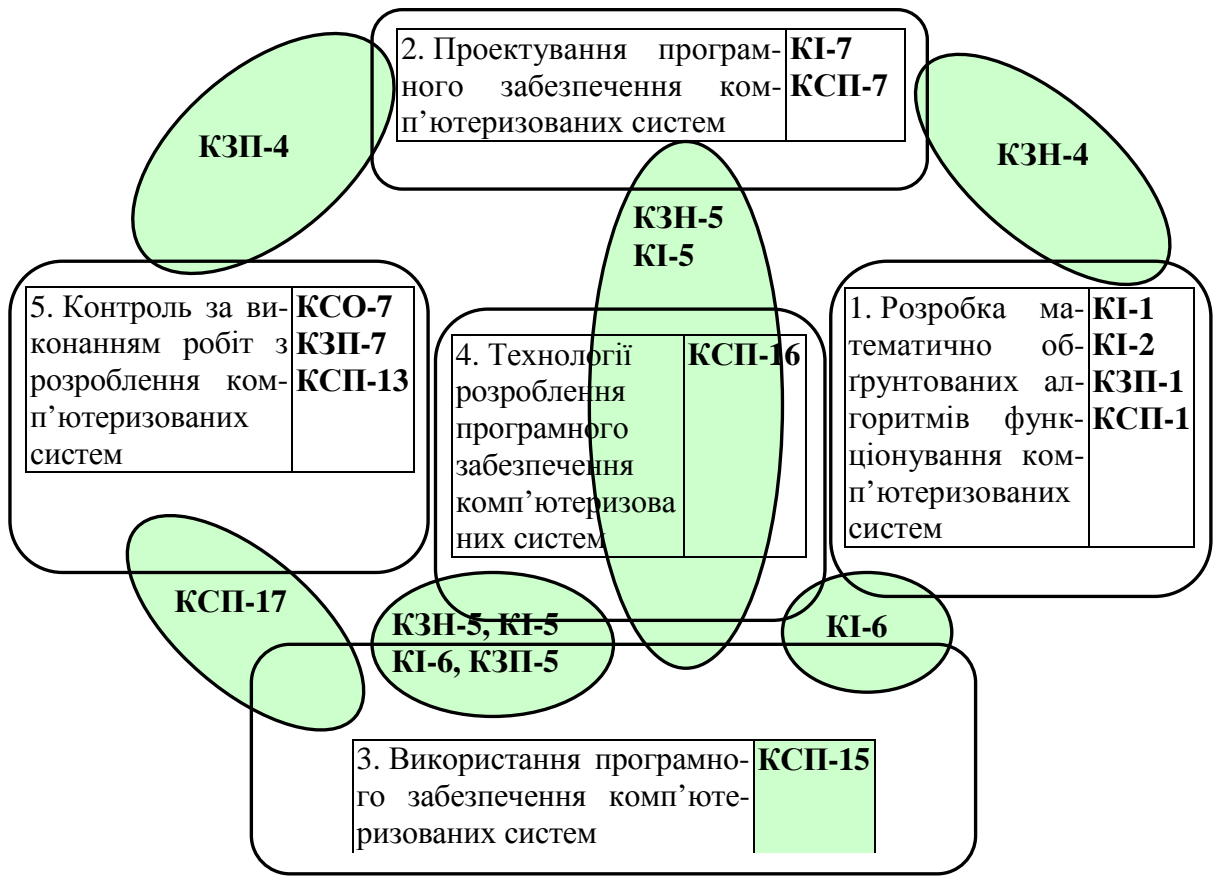

*Рис. 2. Структура компетентності бакалавра інформатики з програмування*

Показана на рис. 2 структура компетентності бакалавра інформатики з програмування надає можливість визначити внесок кожної компетенції, виходячи з кількості зв'язків між типовими задачами діяльності (рис. 3).

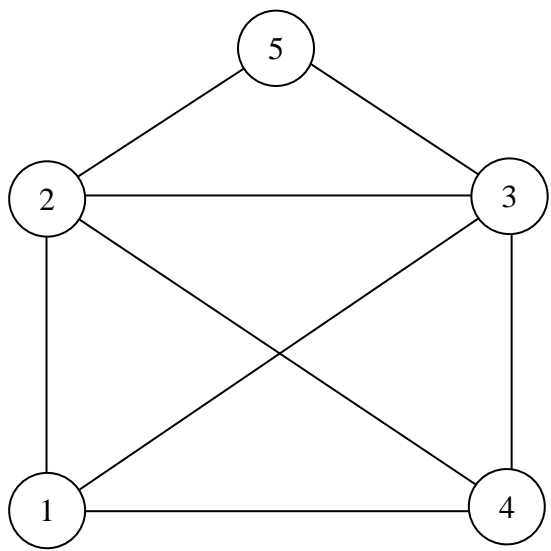

*Рис. 3. Структура компетентності бакалавра інформатики з програмування з виділеними зв'язками*

Конвертоподібна структура, подана на рис. 3, надає можливість зробити висновок про те, що компетентність бакалавра інформатики з програмування є системою. Наразі перші чотири задачі діяльності утворюють повнозв'язну структуру, а п'ята задача є пов'язаною лише з другою і третьою. У табл. 3 показано внесок кожної компетенції у формування компетентності бакалавра інформатики з програмування, визначений пропорційно до кількості задач діяльності, забезпечуваних компетенцією.

*Таблиця 3*

**Внесок компетенцій у формування компетентності бакалавра інформатики з програмування**

| Компетенція | Внесок |
|---|---|
| *Компетенції соціально-особистісні* | *3,57 %* |
| КСО-7. Знання та розуміння законів та методів міжособистісних комунікацій, норм толерантності, ділових комунікацій у професійній сфері, ефективної праці в колективі, адаптивності | 3,57 % |
| *Загальнонаукові компетенції* | *17,86 %* |
| КЗН-4. Базові знання із системних та кібернетичних наук, необхідних для засвоєння загально-професійних дисциплін з інформатики | 7,14 % |
| КЗН-5. Базові знання в галузі інформатики й сучасних інформаційних технологій | 10,71 % |
| *Інструментальні компетенції* | *32,14 %* |
| КІ-1. Знання та розуміння правил письмової й усної рідної мови | 3,57 % |
| КІ-2. Знання та розуміння правил письмової й усної іноземної мови (мов) | 3,57 % |
| КІ-5. Знання методів та правил роботи з комп'ютером та роботи в Інтернеті | 10,71 % |
| КІ-6. Знання законів, методів та методик проведення наукових та прикладних досліджень | 10,71 % |
| КІ-7. Володіння навичками користувача офісних технологій у контексті опрацювання економічної інформації | 3,57 % |
| *Загально-професійні компетенції* | *21,43 %* |
| КЗП-1. Знання методології системних досліджень, методів дослідження та аналізу складних природних, техногенних, економічних та соціальних об'єктів | 3,57 % |

| Компетенція | Внесок |
|---|---|
| та процесів, розуміння складності об'єктів та процесів різної природи, їх різноманіття, багатофункціональність, взаємодію та умови існування для розв'язання прикладних і наукових завдань в галузі системних наук та кібернетики | |
| КЗП-4. Знання вимог чинних державних та міжнародних стандартів, методів і засобів проектування комп'ютеризованих систем, життєвого циклу їх програмного забезпечення | 7,14 % |
| КЗП-5. Знання та розуміння основ програмування, мов різних рівнів та їхніх переваг для розв'язання конкретних задач, методів розроблення програмного забезпечення комп'ютеризованих систем з використанням сучасних технологій | 7,14 % |
| КЗП-7. Знання основних методів та підходів щодо організації, планування, керування та контролю роботами з проектування, розроблення, післяпроектного супроводу та експлуатації програмного забезпечення комп'ютеризованих систем | 3,57 % |
| *Спеціалізовано-професійні компетенції* | *25 %* |
| КСП-1. Знання та розуміння методів системного аналізу та теоретичної кібернетики щодо побудови інформаційних моделей об'єктів та процесів різної природи | 3,57 % |
| КСП-7. Знання основних парадигм проектування та мов моделювання програмного забезпечення комп'ютеризованих систем, методів планування життєвого циклу програмного забезпечення та розроблення моделі керування ресурсами | 3,57 % |
| КСП-13. Знання методів, нормативів, державних стандартів та чинного законодавства стосовно організації, планування, контролю та управління роботами з проектування та розроблення комп'ютеризованих систем колективом розробників | 3,57 % |
| КСП-15. Знання операційних систем (Windows, Unix тощо), системного програмного забезпечення, найбільш розповсюджених пакетів прикладних програм, інформаційних порталів Інтернет, програмних методів захисту інформації в комп'ютеризованих системах та мережах | 3,57 % |
| КСП-16. Знання базових та спеціалізованих технологій розроблення програмного забезпечення комп'ютеризованих систем | 3,57 % |
| КСП-17. Знання методів, методик контролю та тестування правильності роботи програмного забезпечення комп'ютеризованих систем | 7,14 % |

На рис. 4 показано співвідношення блоків компетенцій, що складають компетентність бакалавра інформатики з програмування. З рисунка видно, що внесок блоку соціально-особистісних компетенцій у компетентність бакалавра інформатики з програмування — найменший (4%). Ці компетенції відносяться лише до п'ятої типової задачі діяльності бакалавра інформатики (контроль за виконанням робіт з розроблення комп'ютеризованих систем), яка є слабко пов'язаною з іншими чотирма типовими задачами діяльності.

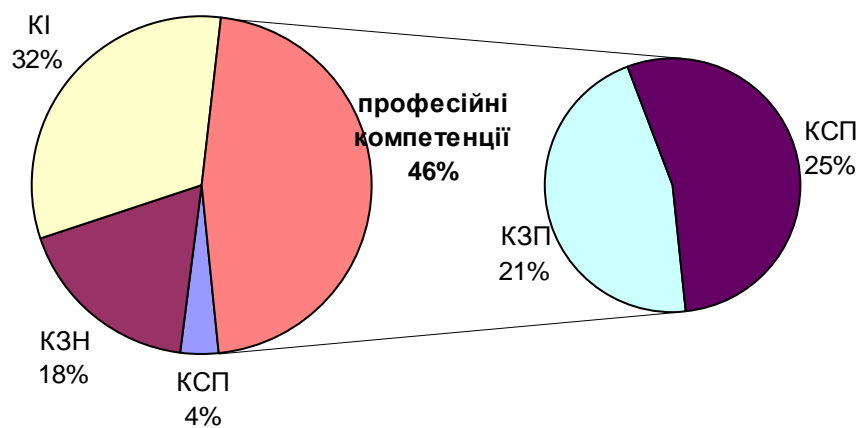

*Рис. 4. Професійні компетенції в компетентності бакалавра інформатики з програмування*

## 3. ВИСНОВКИ ТА ПЕРСПЕКТИВИ ПОДАЛЬШИХ ДОСЛІДЖЕНЬ

У результаті проведеного теоретичного дослідження встановлено, що формування компетентності бакалавра інформатики з програмування вимагає залучення змістових модулів, що відносяться не лише до спеціальних (професійно орієнтованих) дисциплін. Суттєвим є внесок у формування компетентності бакалавра інформатики з програмування компетенцій, що традиційно не виокремлюються у структурі компетентності з програмування, зокрема: соціально-поведінкові, комунікативні, загальнонаукові, управлінські та дослідницькі (разом — 54 %). Це створює умови для розгляду такої компетентності як фундаментальної і надає можливість її ефективного формування у різних середовищах навчання, провідними з яких є хмаро орієнтовані.

## СПИСОК ВИКОРИСТАНИХ ДЖЕРЕЛ

# КОМПЕТЕНТНОСТЬ БАКАЛАВРА ИНФОРМАТИКИ В ПРОГРАММИРОВАНИИ

**Стрюк Андрей Николаевич**
кандидат педагогических наук, докторант
Институт информационных технологий и средств обучения НАПН Украины, г. Киев, Украина
*andrey.n.stryuk@gmail.com*

**Семериков Сергей Алексеевич**
профессор, доктор педагогических наук, заведующий кафедрой
фундаментальных и социально-гуманитарных дисциплин
ГВУЗ «Криворожский национальный университет», г. Кривой Рог, Украина
*semerikov@gmail.com*

**Тарасов Игорь Владимирович**
ассистент кафедры информатики и прикладной математики
ГВУЗ «Криворожский национальный университет», г. Кривой Рог, Украина
*taras_2001@rambler.ru*

**Аннотация.** На основе анализа подходов к определению профессиональных компетентностей специалистов в области информационных технологий выделена компетентность бакалавра информатики в программировании. С учетом отраслевого стандарта высшего образования по направлению подготовки 040302 «Информатика» и Computing Curricula 2001 определены содержание и структура компетентности бакалавра информатики в программировании. Спроектирована система содержательных модулей, обеспечивающих ее формирования. Определен вклад нормативных компетенций бакалавра информатики в формирование компетентности в программировании. Предложены направления формирования компетентности в программировании, в частности, в облачно ориентированной среде обучения.

**Ключевые слова:** компетентность в программировании; подготовка бакалавров информатики; профессиональные компетенции бакалавра информатики.

# BACHELOR OF INFORMATICS COMPETENCE IN PROGRAMMING

**Andrii M. Striuk**
Ph. D. (pedagogical sciences), doctorate student
Institute of Information Technologies and Learning Tools of NAES of Ukraine, Kyiv, Ukraine
*andrey.n.stryuk@gmail.com*

**Serhiy O. Semerikov**
Professor, D. Sc. (pedagogical sciences), head of the Department fundamental and sociohumanitarian disciplines
SIHE "Kryvyi Rih National University", Kryvyi Rih, Ukraine
*semerikov@gmail.com*

**Ihor V. Tarasov**
Assistant Professor of the Department of informatics and applied mathematics
SIHE "Kryvyi Rih National University", Kryvyi Rih, Ukraine
*taras_2001@rambler.ru*

**Abstract.** Based on the analysis of approaches to the definition of professional competencies of IT students the competence in programming of bachelor of informatics is proposed. Due to the standard of training in 040302 "Informatics" and Computing Curricula 2001 it was defined the content and structure of the competence in programming of bachelor of informatics. The system of content modules providing its formation was designed. The contribution of regulatory competencies of bachelor of informatics in the formation of competence in programming is defined. The directions of formation of competence in programming in the cloudy-oriented learning environment are proposed.

**Keywords:** competence in programming; training of bachelors of informatics; professional competencies of bachelor of informatics.

## REFERENCES (TRANSLATED AND TRANSLITERATED)